\documentclass[twocolumn]{aastex61}
\graphicspath{ {Figures/} }
\received{\today}
\revised{\today}
\accepted{\today}
\submitjournal{ApJL}

\shorttitle{Limitations of optical diagnostics in confirming IMBHs}
\shortauthors{J.M.Cann et al}

\begin{document}

\title{The Limitations of Optical Spectroscopic Diagnostics in Identifying AGNs in the Low Mass Regime}

\correspondingauthor{Jenna M. Cann}
\email{jcann@masonlive.gmu.edu}

\author{Jenna M. Cann}
\affiliation{George Mason University, Department of Physics and Astronomy, MS3F3, 4400 University Drive, Fairfax, VA 22030, USA}
\affiliation{National Science Foundation, Graduate Research Fellow}

\author{Shobita Satyapal}
\affiliation{George Mason University, Department of Physics and Astronomy, MS3F3, 4400 University Drive, Fairfax, VA 22030, USA}

\author{Nicholas P. Abel}
\affiliation{MCGP Department, University of Cincinnati, Clermont College, Batavia, OH 45103, USA}

\author{Laura Blecha}
\affiliation{University of Florida, Department of Physics, P.O. Box 118440, Gainesville, FL 32611-8440}

\author{Richard F. Mushotzky}
\affiliation{Department of Astronomy, University of Maryland, College Park, MD 20742-2421, USA}

\author{Christopher S. Reynolds}
\affiliation{Institute of Astronomy, Madingley Road, Cambridge CB3 0HA}

\author{Nathan J. Secrest}
\affiliation{U.S. Naval Observatory, 3450 Massachusetts Avenue NW, Washington, DC 20392, USA}



\begin{abstract}
Intermediate-mass black holes (IMBHs) with masses between $100 - 10^5M_{\odot}$ are crucial to our understanding of black hole seed formation and are the prime targets for {\it LISA}, yet black holes in this mass range have eluded detection by traditional optical spectroscopic surveys aimed at finding active galactic nuclei (AGNs). In this paper, we have modeled for the first time the dependence of the optical narrow emission line strengths on the black hole mass of accreting AGN over the range of $100-10^8M_{\odot}$. We show that as the black hole mass decreases, the hardening of the spectral energy distribution from the accretion disk changes the ionization structure of the nebula. The enhanced high energy emission from IMBHs results in a more extended partially ionized zone compared with models for higher mass black holes. This effect produces a net decrease in the predicted [OIII]/H$\beta$ and [NII]/H$\alpha$ emission line ratios. Based on this model, we demonstrate that the standard optical narrow emission line diagnostics used to identify massive black holes fail when black hole mass falls below $\approx10^4M_{\odot}$ for highly accreting IMBHs and for radiatively inefficient IMBHs with active star formation. Our models call into question the ability of common optical spectroscopic diagnostics to confirm AGN candidates in dwarf galaxies, and indicate that the low-mass black hole occupation fraction inferred from such diagnostics will be severely biased.
\end{abstract}




\section{Introduction} \label{sec:intro}

It is now well-established  that supermassive black holes (SMBH), over a million times the mass of the sun, exist in the centers of galaxies and that their masses are well-correlated with properties of their host galaxies \citep[e.g.,][]{magorrian1998, gebhardt2000, gultekin2009, mcconnell2013}. In contrast, the existence, properties, and host galaxy demographics of black holes with masses between a few tens of solar masses and under a million solar masses are not well-understood. This is a significant deficiency, since the study of this population could provide insight into the origin and evolution of SMBH seeds, thought to have formed at high redshift.  The mass function and occupation fraction of these ``intermediate-mass black holes (IMBHs)" in the local universe provide crucial constraints to models of seed formation, potentially allowing us to discriminate between lower mass seeds formed from stellar remnants or massive seeds formed by the direct collapse of dense gas \citep[see][and references therein]{natarajan2014}. Moreover, mergers between IMBHs are one of the most promising sources of gravitational waves (GWs) detected with the Laser Interferometer Space Antenna \citep[LISA;][]{amaroseoane2013}, however no IMBH pairs have been identified and their merger rate is unknown.  Finding a population of IMBHs, measuring their masses, determining their merger rates, and understanding their connection to their host galaxies is therefore of fundamental astrophysical importance.

Unfortunately, due to their low mass, IMBHs can only be detected in a significant sample of galaxies if they are accreting \citep[e.g.,][]{greene2012}. One of the most widely used methods to identify accreting black holes or active galactic nuclei (AGNs) employs prominent optical spectroscopic emission line ratios \citep{baldwin1981, VeilleuxOsterbrock1987, kewley2001} in which AGNs and HII regions typically separate in two-dimensional line-diagnostic diagrams. This method exploits the fact that emission line ratios will vary in response to the radiation field responsible for ionizing the gas, which can be either entirely stellar in origin or dominated by a centrally located hard radiation field produced by an accretion disk around a massive black hole. The most widely used diagnostic diagram uses the [OIII]/H$\beta$ versus the [NII]/H$\alpha$ emission line ratios (commonly referred to as the Baldwin-Phillips-Terlevich, BPT, diagram), in which most known AGNs exhibit higher line ratios than star-forming galaxies \citep{kewley2001}. Using these diagnostics, there have been a growing number of AGNs discovered in low mass galaxies or galaxies lacking classical bulges \citep[e.g.,][]{reines2013}. However, despite the vast amount of optical spectroscopic data available from the Sloan Digital Sky Survey, to date, there exist only a small fraction of dwarf galaxies optically classified as hosting AGNs. Indeed, a key and striking result based on optical spectroscopic studies, is that the fraction of galaxies with signs of accretion activity drops dramatically at stellar masses  $\log{M_\star/M_\sun}<10$ \citep[e.g.][]{kauffmann2003}. In fact, for a sample of dwarf galaxies with stellar masses $\log{M_\star/M_\sun}<9.5$ and high quality optical emission line measurements, only $0.1\%$ of galaxies are unambiguously identified as AGNs based on their emission line ratios \citep{reines2013}, compared to $>80\%$ of galaxies with $\log{M/M_{\odot}}>11$ \citep{kauffmann2003}.  While this suggests AGN activity is less prevalent in low mass galaxies, the effectiveness of optical emission line ratios in identifying accreting \textit{low mass} black holes has not been established.  

It is well known that optical spectroscopic diagnostics can fail at identifying AGNs in galaxies with active star formation, where photoionization from stars and starburst-driven winds can dominate the optical spectrum, and gas and dust can obscure the central engine \citep[e.g.][]{goulding2009,kewley2013,trump2015}.  However, it is not known how these emission line ratios depend on the mass of the black hole. This is of prime importance for high spatial resolution follow-up observations, where it has generally been assumed that if contamination from surrounding star formation is reduced, the AGN can be identified.  Indeed, follow-up optical spectroscopic observations using high spatial resolution integral field units (IFUs) are currently the gold standard to confirm or refute an AGN candidate identified through multiwavelength studies. Many more AGN candidates are expected to be identified at X-ray wavelengths with the launch of eRosita in 2019 \citep{erosita}.  Recent work by Agostino \& Salim (2018, submitted to ApJ) demonstrates that there is a significant population of X-ray identified AGN that have BPT line ratios consistent with star-forming galaxies. With a limiting flux of $\approx 10^{-13}~erg/cm^2/s$ in the 2-10 keV band, the detection of black holes with masses as low as $1000M_{\odot}$ is possible within 10 Mpc.  As the SDSS catalog has $\approx3000$ high S/N dwarf galaxies with masses less than $10^8M_{\odot}$, and $\approx1200$ with masses less than $10^7M_{\odot}$, there will be a large sample available in which optical spectra can be used to constrain X-ray contribution from stellar processes. Reliance on optical spectroscopic confirmation could therefore severely bias conclusions about the black hole occupation fraction.  As the optical spectroscopic  diagnostic diagrams have been established based on semi-empirical classification schemes using galaxy samples with black hole masses in excess of $10^6M_{\odot}$ \citep{kewley2001,kauffmann2003} and photoionization models using a stellar ionizing continuum and a single power law to model the ionizing radiation field of the AGN  \citep{VeilleuxOsterbrock1987,kewley2001}, these models do not take into account the effect of black hole mass on the ionizing radiation field, which in turn would impact the predicted emission line spectrum and potentially the currently employed optical BPT AGN classification schemes for IMBHs.  As the black hole mass decreases, the Schwarzchild radius decreases, and in response, the temperature of the surrounding accretion disk increases.  The shape of the ionizing radiation field therefore changes with black hole mass, which in turn will impact the emission line spectrum at optical wavelengths, potentially affecting the location on the BPT diagram.

In this paper, we explore the dependence of the BPT emission line ratios as a result of changes in black hole mass.  The goal of our work is to determine if the standard optical spectroscopic diagnostics used widely to identify AGNs can be applied to accreting IMBHs.  In Section 2, we describe our model and discuss our photoionization calculations.  In Section 3, we show our calculated BPT line ratios as a function of black hole mass.  We discuss the implications of these results in Section 4 and our conclusions in Section 5.

\section{Theoretical Calculations} \label{sec:theory}

In this paper, we model the optical emission line ratios with a simple photoionization model assuming the ionizing radiation field is produced exclusively by an AGN continuum.  The details of this model, using XSPEC \citep{arnaud96} and Cloudy c17 \citep{ferland2017}, are discussed in \citet{cann2018}. Briefly, the AGN is modeled as a simple geometrically thin, optically thick Shakura-Sunyaev \citep{shakura1973} accretion disk with Comptonized X-ray radiation in the form of a power law and a soft-excess component. The accretion disk temperature changes as a function of black hole mass as given by:

\begin{equation}
T=6.3\times10^{5}\bigg(\frac{\dot{m}}{\dot{m}_{Edd}}\bigg)^{1/4}\bigg(\frac{M_{BH}}{10^{8}M_{\odot}}\bigg)^{-1/4}\bigg(\frac{R}{R_s}\bigg)^{-3/4}K
\end{equation}

We assume a one-dimensional, ionization-bounded spherical model with a closed geometry, where the cloud is between the observer and the continuum source, and the ionization parameter, gas density, and Eddington ratio are allowed to vary. The ionization parameter, $U$, is defined as the dimensionless ratio of the incident ionizing photon density to the hydrogen density:

\begin{equation}
U=\frac{\phi_H}{n_Hc}=\frac{Q(H)}{4\pi R^{2}n_{H}c}
\end{equation}

We assume gas and dust abundances consistent with the local interstellar medium (ISM), and consider solar and 0.1 solar metallicity models. We did not include the effects of shocks.  We point the reader to \citet{cann2018} for details on model parameters and assumptions adopted.  We computed a total of $5,070$ models, where ionization parameter was varied between $\log{U}=-1$ to $\log{U}=-4$ in increments of 0.25 dex, the hydrogen gas density between $\log{n_\mathrm{H}/cm^{-3}}=1.5$ to $3.5$ in increments of 1.0 dex, the Eddington ratio between $10^{-4}-1$ in increments of 1.0 dex, and mass between $100-10^8M_{\odot}$ in increments of 0.5 dex.

\section{Results} \label{sec:results}
\begin{figure}
\includegraphics[width=0.5\textwidth]{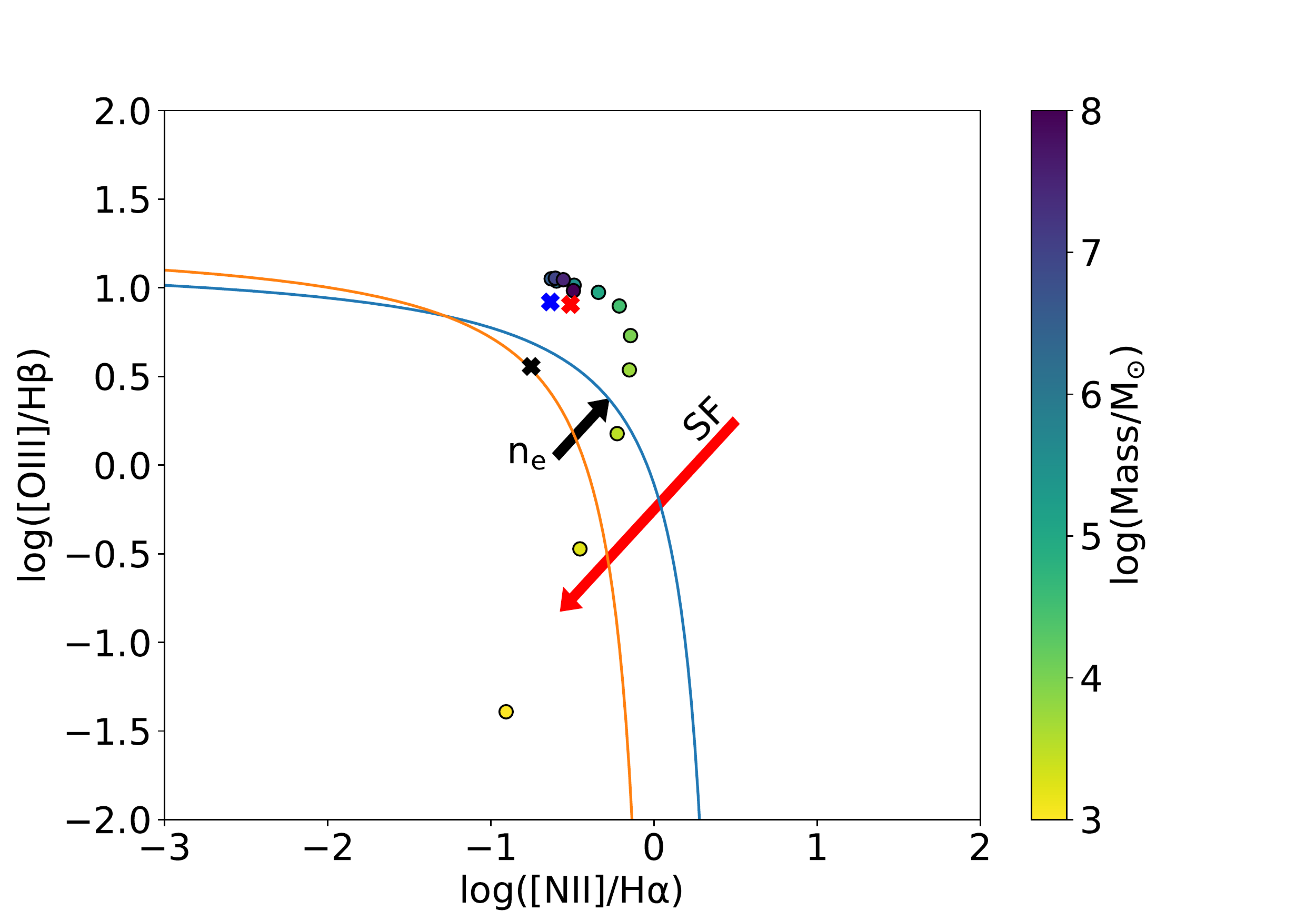}
\includegraphics[width=0.5\textwidth]{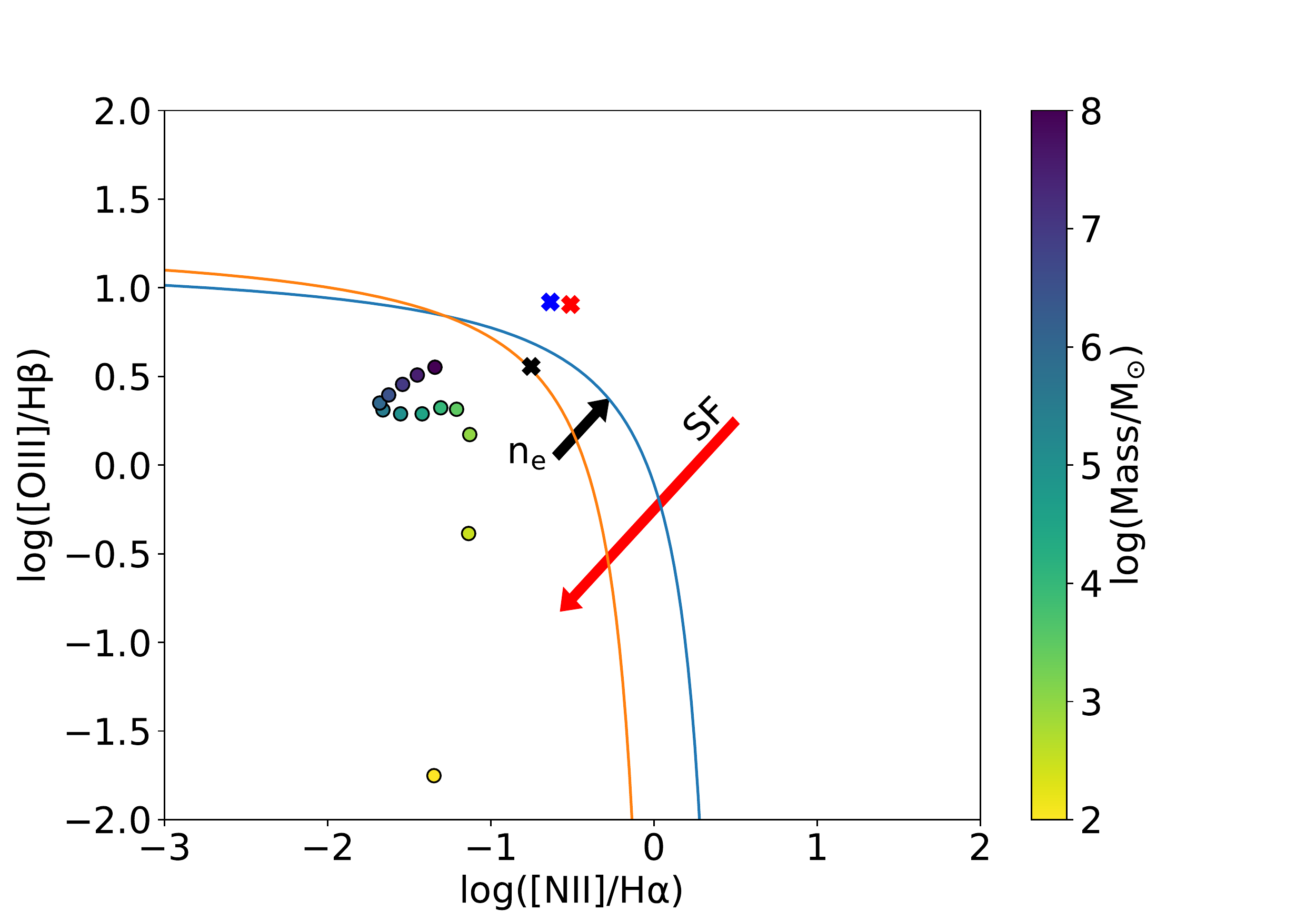}
\caption{BPT diagram for our photoionization models for a range of black hole masses from $10^3-10^8M_{\odot}$ at constant $\log{U}=-2$, $n_H=300~cm^{-3}$, and $\dot{m}/\dot{m}_{Edd}=0.1$ at solar (top) and 0.1-solar (bottom) metallicities. The red and blue lines correspond to demarcations separating star-forming galaxies from AGNs from \citet{kauffmann2003} (red) and \citet{kewley2001} (blue), respectively.  For black hole masses $<1000 M_{\odot}$ in the top panel, the line ratios fall outside the plotted region shown in the figure. We also show the observed BPT line ratios of RGG118 (black 'x') reported by \citet{baldassare2017}, POX 52 (red 'x') reported by \citet{barth2004}, and NGC 4395 (blue 'x') reported by \citet{kraemer1999}.  Note our models do not include star formation.  The arrows denote the direction that model points would move in the presence of star formation (red) and higher electron densities (black).}
\label{fig1}
\end{figure}

In Figure \ref{fig1}, we show the optical BPT diagram as a function of black hole mass for the $\dot{m}=0.1$, $n_H=300~cm^{-3}$, $\log{U}=-2.0$ model.  Note that, typically, $U$ is $\sim10^{-3}$ based on observations of optical emission lines in star-forming galaxies and HII regions \citep{dopita2000, moustakas2010}, but values as high as $\log{U}=-2.0$ are found in regions such as ULIRGs \citep{abel09} or high redshift galaxies \citep[e.g.,][]{brinchmann08,erb2010}. For illustrative purposes, we plot in Figure \ref{fig1} the higher ionization parameter case, which may be more typical in dwarf galaxies \citep{izotov2001}. The effect of ionization parameter on the line ratios is shown in Figure \ref{fig3}.  We also plot the widely adopted AGN demarcation lines used in the literature to identify AGNs, and the location of RGG118 \citep{baldassare2017}, a dwarf galaxy recently found to host a $50,000M_{\odot}$ black hole, the lowest mass SMBH currently known.  As can be seen, our model predicts that as black hole mass decreases,  the line ratios fall outside the widely used AGN and composite demarcation regions of the diagram, and that the transition mass is at approximately $10,000M_{\odot}$. Interestingly, the line ratios of RGG118 fall right on the edge of the star-forming/composite demarcation line, consistent with the predictions of our models given its black hole mass. We also show the effects of changes in gas density, metallicity, and adding a contribution from star formation. Note that both lowering the metallicity and increasing the contribution from star formation, both of which likely accompany IMBH hosts, results in line ratios that move further into the star-forming region of the diagram.

\begin{figure}
\includegraphics[width=0.495\textwidth]{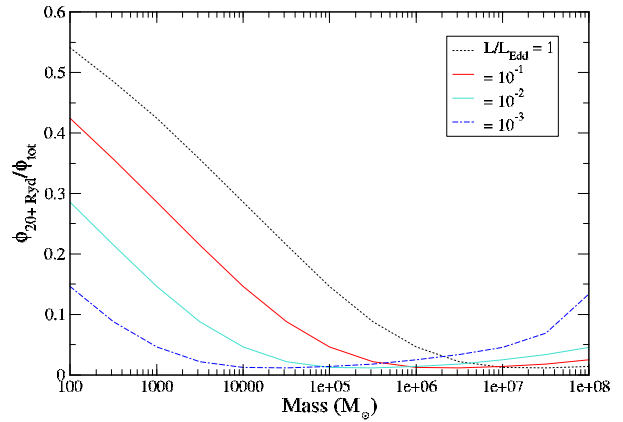}
\caption{Fraction of total flux in X-ray photons with energies greater than $20$ Ryd striking the illuminated face of the cloud per second compared to total photon flux per second as a function of black hole mass for a range of Eddington ratios.  As can be seen, in low mass black holes, a large fraction of radiation emitted is over $20$ Ryd.}
\label{fig2}
\end{figure}

\begin{figure}
\includegraphics[width=0.492\textwidth]{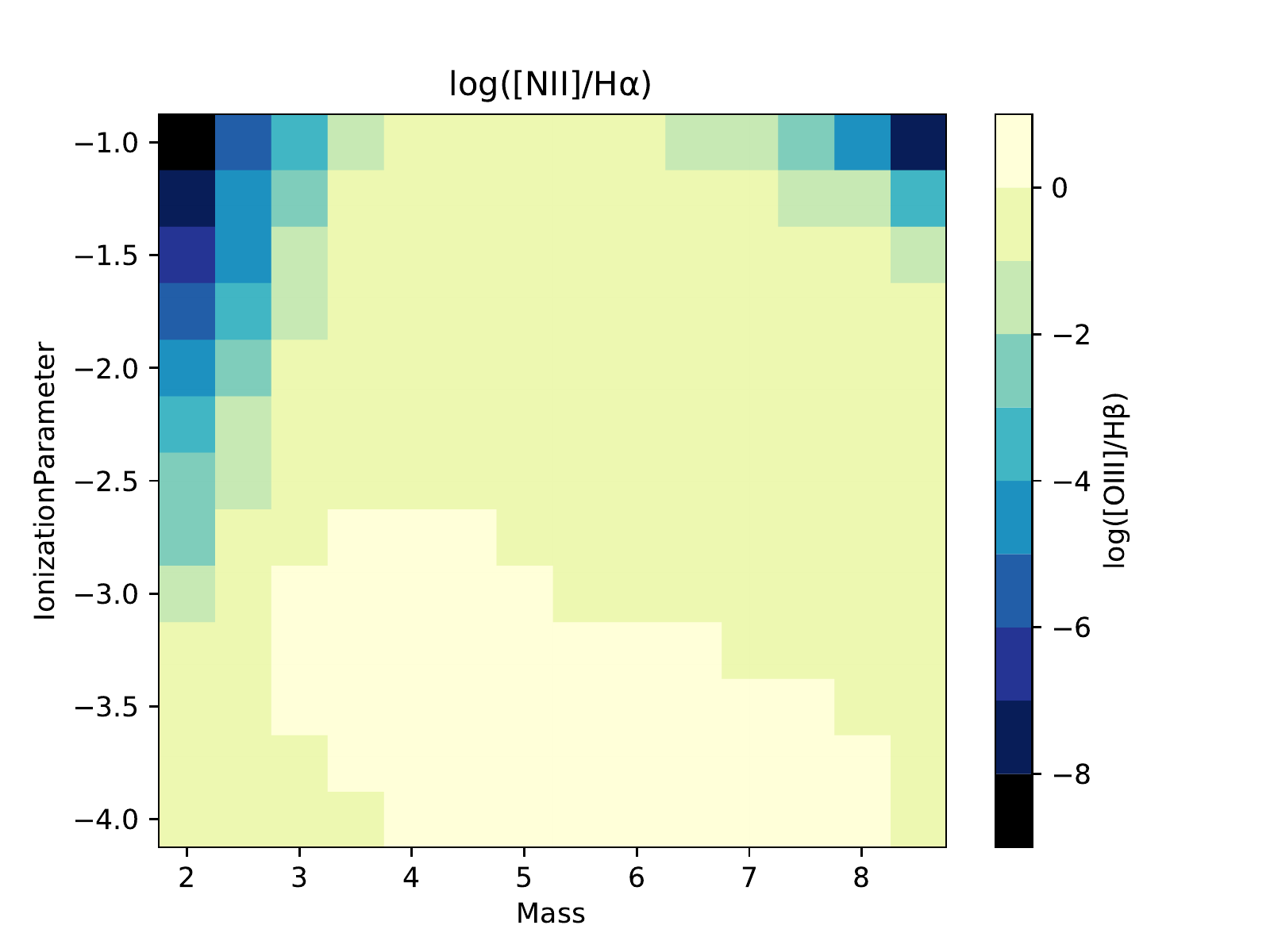}
\includegraphics[width=0.492\textwidth]{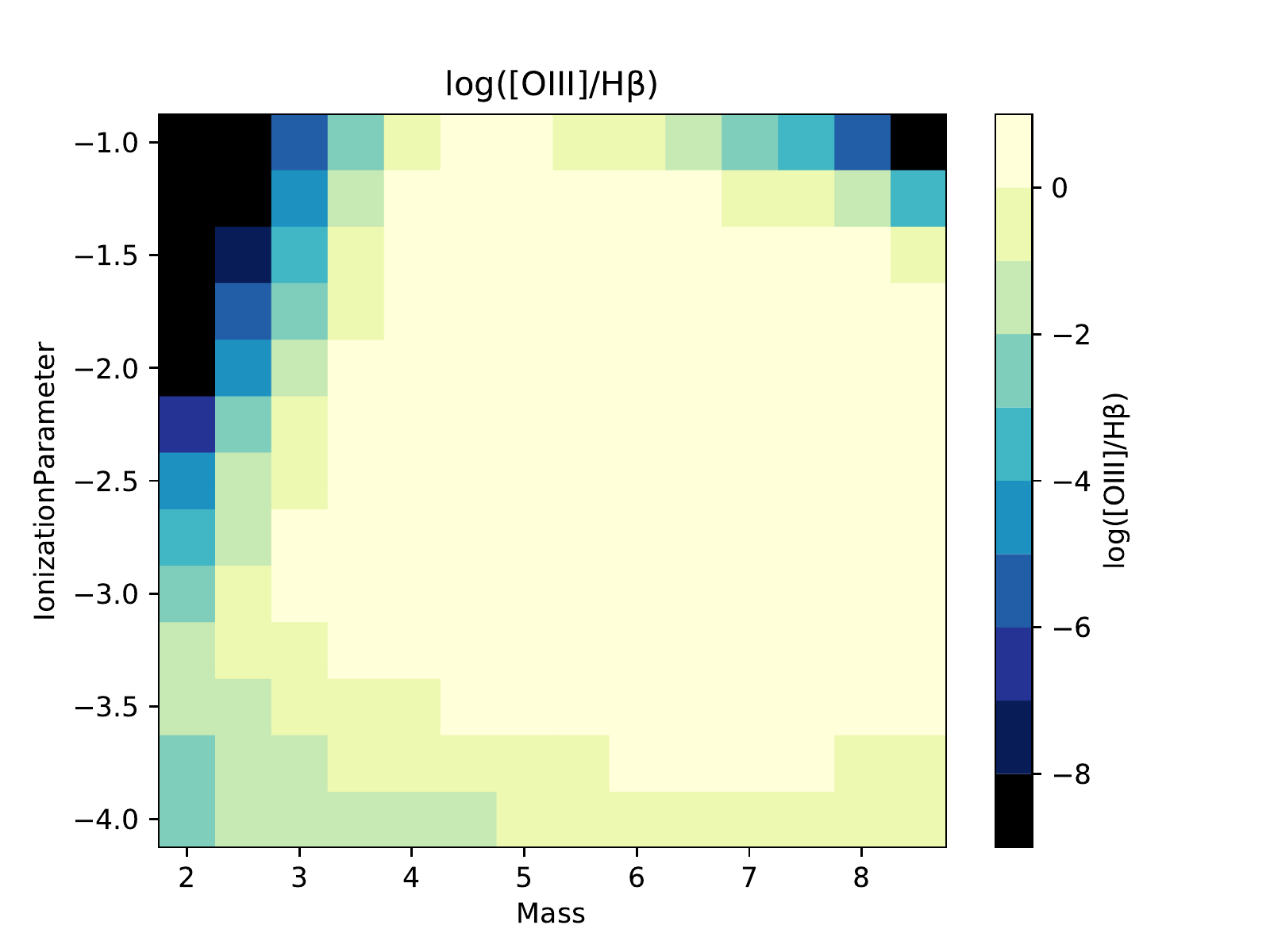}
\caption{Contour plots showing the changes of the BPT line ratios over a range of BH mass from $100 - 10^8$ M$_{\odot}$ and ionization parameter from $\log$U = $-1$ to $-4$ for $\dot{m}/\dot{m}_{Edd} = 0.1$ and $\log{n_H} = 300$ cm$^{-3}$.}
\label{fig3}
\end{figure}

The observed behavior of the line ratios with black hole mass is a consequence of two main factors. As the black hole mass decreases, the resulting hardening of the AGN spectral energy distribution (SED) changes the ionization structure of the nebula. For massive black holes in the $10^7-10^8M_{\odot}$ range, the dominant ionization states of oxygen are $O^{+}$ and $O^{2+}$, but as the black hole mass decreases, and the ionizing radiation field shifts to higher energies, some of the O is found in higher ionization states, extending even up to $O^{8+}$ for the lowest masses modeled. In addition, as the black hole mass decreases, the enhanced X-ray emission from the accretion disk penetrates further into the cloud, resulting in a significantly extended partially ionized zone where $H^{+}$ is produced but $O^{2+}$ is not. This effect results in a net decrease in the predicted [O~III]/H$\beta$ emission line ratio. The fraction of Nitrogen in $N^{+}$, on the other hand, is relatively constant as a function of black hole mass over the range explored in our models. However, the extended partially ionized zone results in a overall decrease in the [N~II]/H$\alpha$ emission line ratio. Note it is well known that AGNs in general produce a much more extended partially ionized zone in which collisionally excited forbidden lines can be produced than is seen in HII regions around massive young stars. This is because the ionizing radiation field produced by a stellar continuum produces very few X-ray photons, resulting in a much sharper ionization front.  The fraction of X-ray photon flux relative to the total flux as a function of black hole mass is shown in Figure \ref{fig2}. As can be seen, there is a steep increase in the fraction of X-ray photons when the black hole mass falls below $10^5M_{\odot}$.

\begin{figure}
\includegraphics[width=0.492\textwidth]{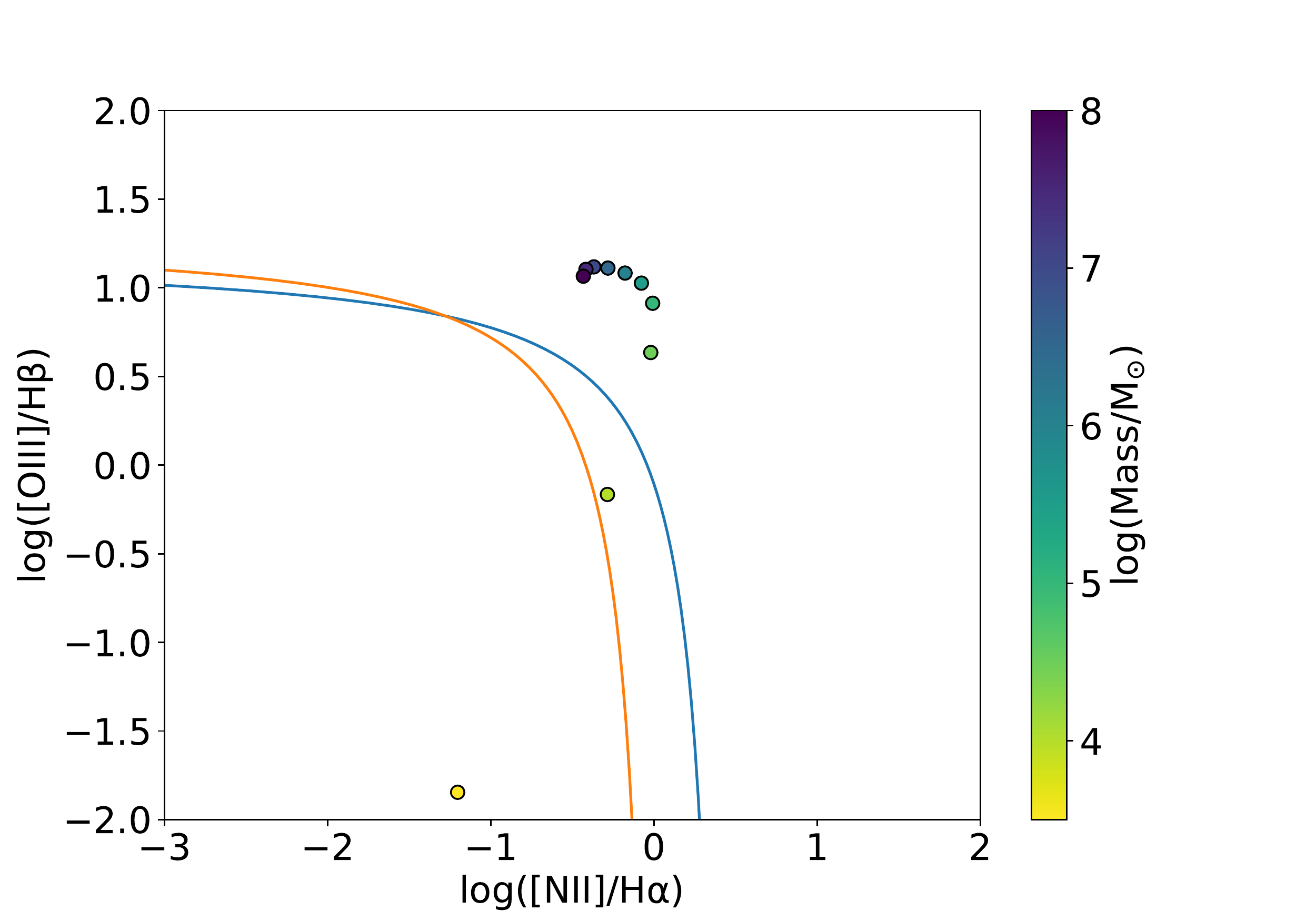}
\includegraphics[width=0.492\textwidth]{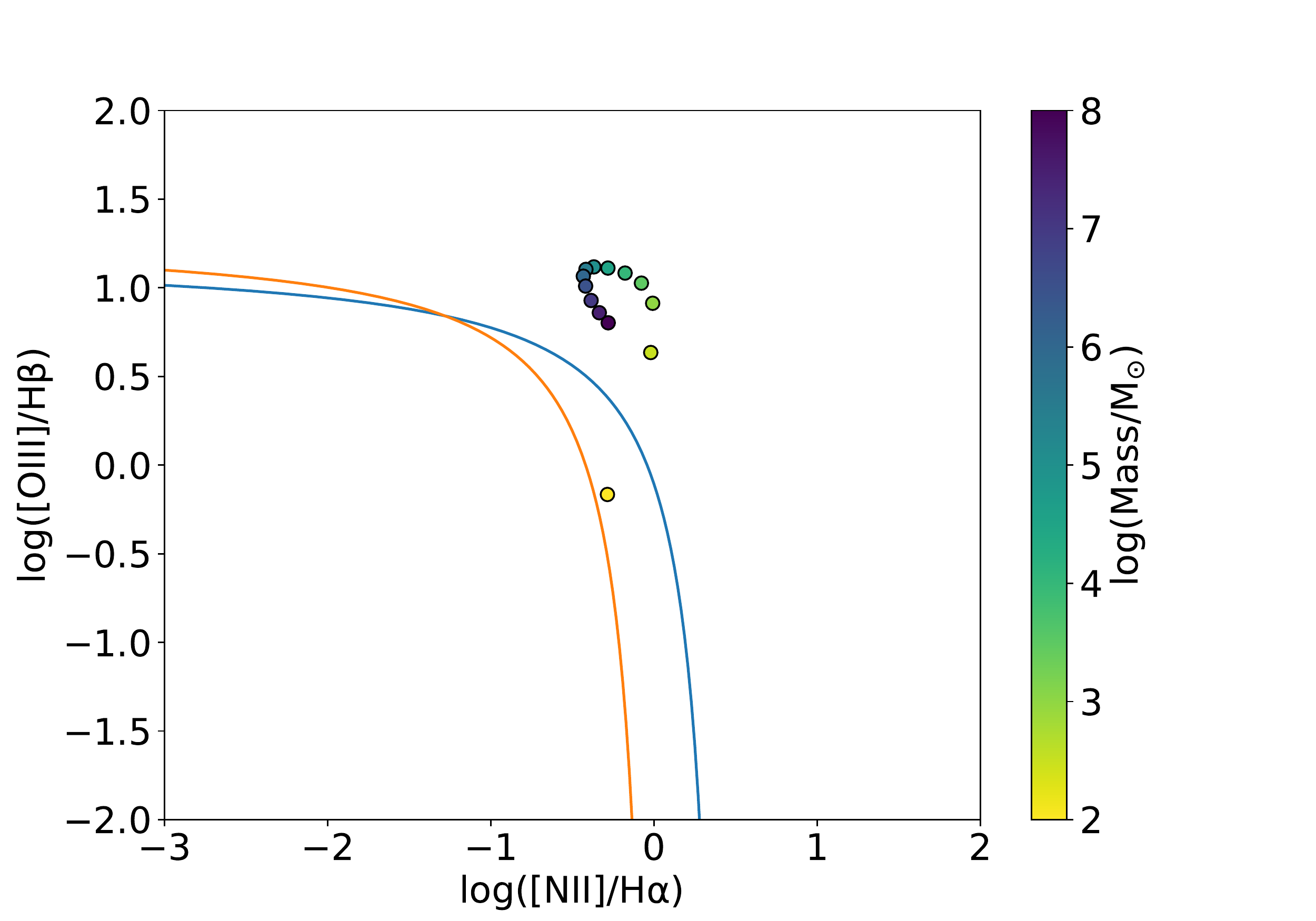}
\caption{BPT plots for $\dot{m}/\dot{m}_{Edd} = 1$ (top) and $0.01$ (bottom) with $\log{U} = -2.5$ and $n_H = 300~cm^{-3}$.  Note that masses not pictured here fall outside the plotted region shown in the figure, below the non-AGN demarcation. Red and blue lines are as denoted in Figure \ref{fig1}. The presence of star formation would move the model points in the direction of the red arrow in Figure \ref{fig1}. Note that as you lower the Eddington ratio, the temperature of the accretion disk decreases such that for the most massive black holes, the resulting softening of the SED of the accretion disk causes a net decrease in the [OIII]/H$\beta$, as seen in the lowest panel above.}
\label{fig4}
\end{figure}

We note that we have shown the effect of black hole mass on the BPT diagram for typical ISM conditions and a single Eddington ratio. The line ratios are of course a strong function of the ionization parameter, since the ionization parameter affects the ionization structure of the nebula. The size of the $H^{+}$ region and the dominant stage of ionization increases with increasing ionization parameter. In Figure \ref{fig3}, we show the effect of both black hole mass and ionization parameter on the BPT line ratios. For typical ionization parameters, which are between $-3.2<\log{U}<-2.9$ for local HII regions \citep{dopita2000} and local star-forming galaxies \citep{moustakas2010}, both the [OIII]/H$\beta$ and the [NII]/H$\alpha$ emission line ratios are lower for IMBHs compared with black holes above $10^6M_{\odot}$, and never make it into the AGN demarcation for black hole masses below $10^3M_{\odot}$. Lower ionization parameter models show a wider range of black hole masses emitting in the AGN regime, but photoionization from star formation will result in line ratios that shift toward the star-forming region of the diagram.  Note the models we have shown here are for a fixed Eddington ratio. The shape of the AGN continuum will change as a function of the Eddington ratio, which in turn will impact the line ratios as seen in Figure \ref{fig4}. As can be seen, the low mass black holes radiating at higher Eddington ratios fall below the AGN demarcation region, but if the Eddington rate is reduced, they move to the traditional AGN regime.

\begin{figure}
\includegraphics[width=0.492\textwidth]{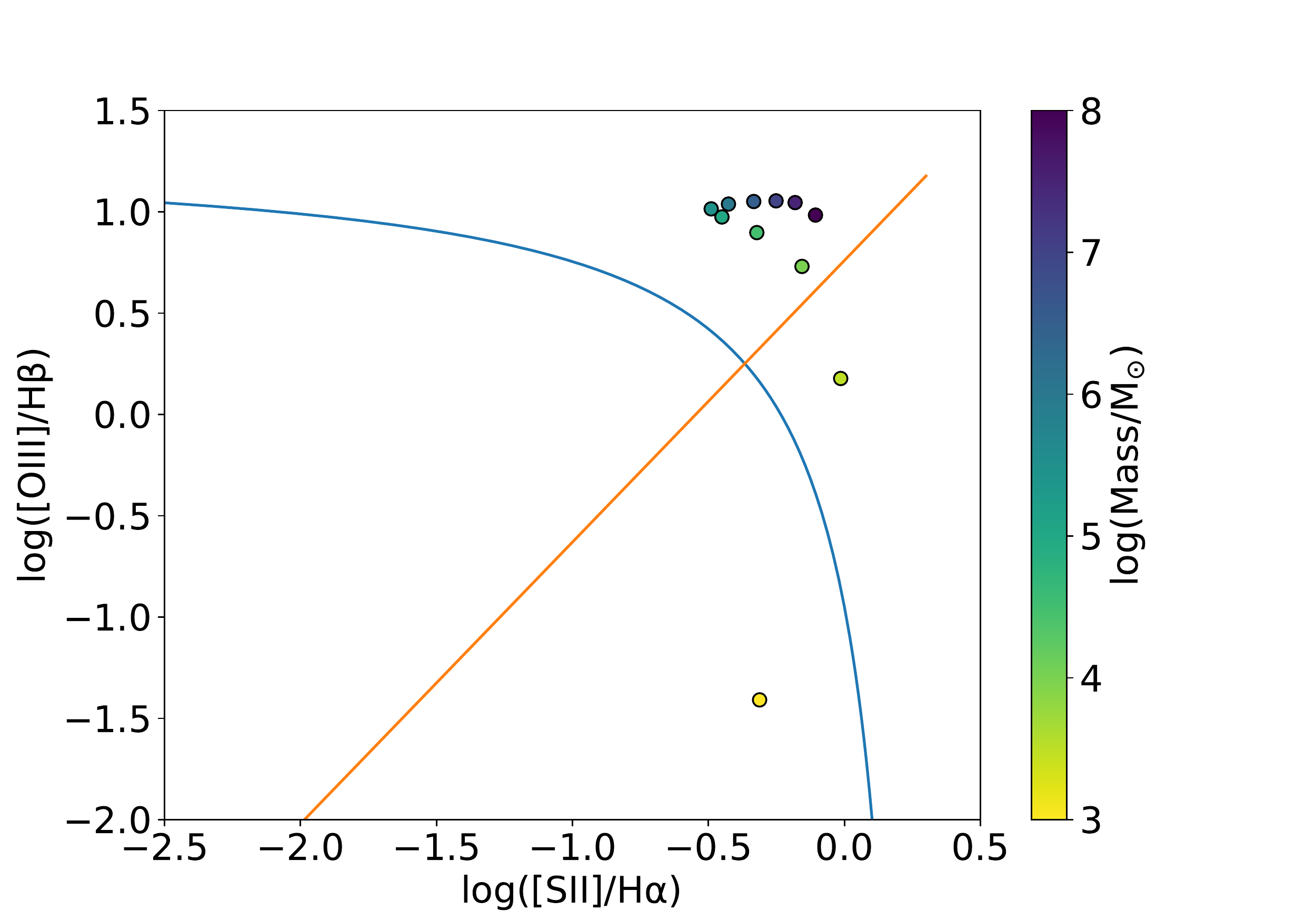}
\includegraphics[width=0.492\textwidth]{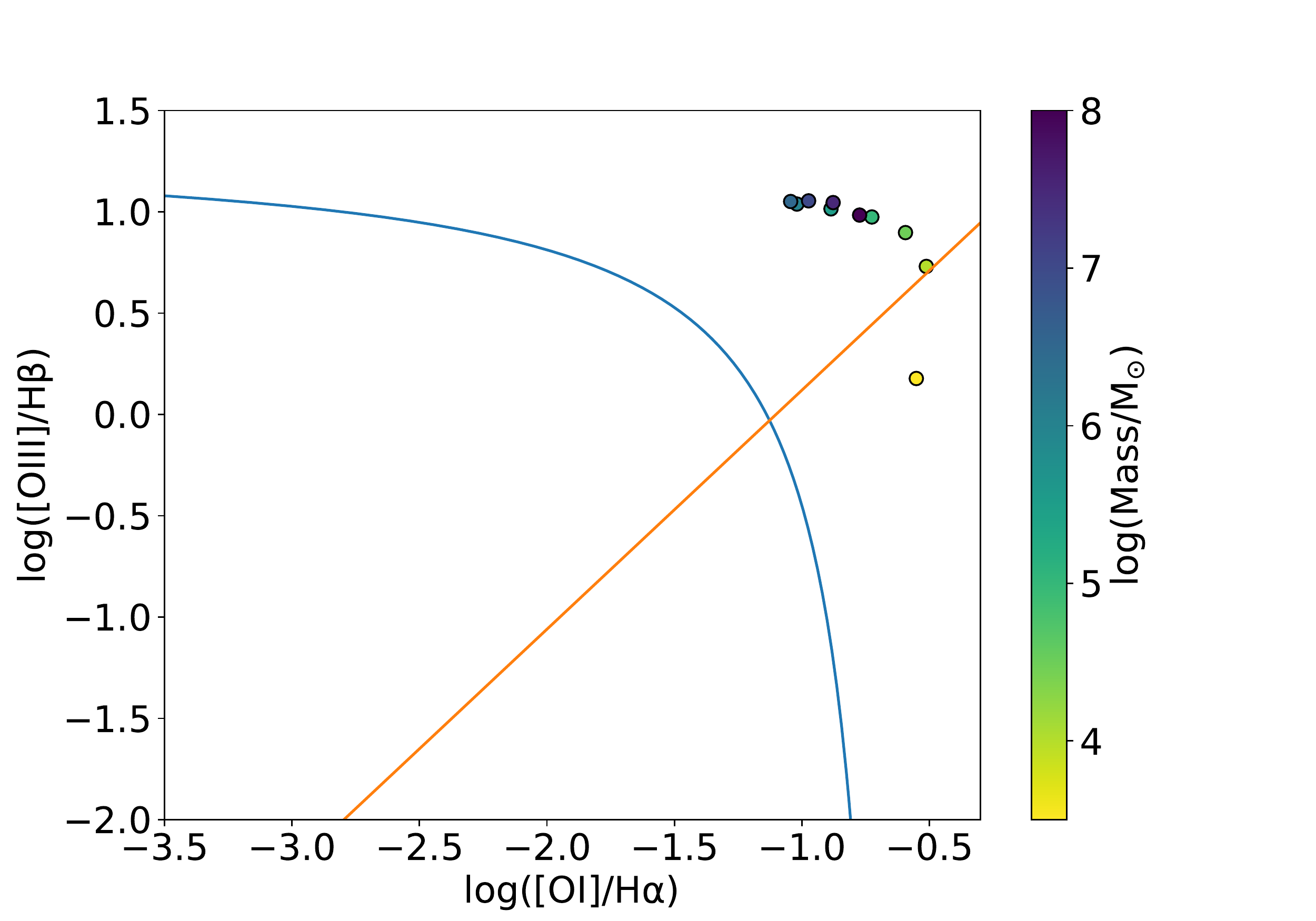}
\caption{Additional BPT diagrams using the [S~II]/H$\alpha$ (top) and [O~I]/H$\alpha$ (bottom) line ratios for the $n_H=300~cm^{-3}$, $\dot{m}/\dot{m}_{Edd}=0.1$, solar metallicity, $\log{U}=-2$ model.  The blue and red lines correspond to the demarcations separating AGN and star-forming galaxies (blue) and AGN from LINERS (red).  Masses below $10^3M_{\odot}$ in the top panel and below $10^{3.5}M_{\odot}$ in the lower panel are outside the range plotted, in the star-forming region.}
\label{fig5}
\end{figure}

There are additional optical diagnostics used to confirm and identify AGN, using the [S~II]/H$\alpha$ and [O~I]/H$\alpha$ line ratios.  The dependence of these ratios on black hole mass was also tested, with the results shown in Figure \ref{fig5}.  As can be seen, a larger range of black hole masses have line ratios that fall in the AGN regime of the graph, however, the line ratios dramatically change and fall into the star-forming region of the plot when black hole mass falls below $10^3-10^{3.5}M_{\odot}$.  If optical diagnostics are to be taken for a candidate IMBH, it is recommended that all BPT line ratios be observed and considered to increase the possibility of an accurate identification, though the lack of AGN colors is still not enough to disregard a strong candidate for the lowest masses.

\section{Discussion} \label{sec:disc}

The models presented in this work call into question the completeness of optical BPT diagrams in confirming the presence of AGNs powered by IMBHs. For a non-negligible region in parameter space, our models predict that low mass AGNs do not produce optical emission line ratios occupied by higher mass black holes. In fact, a $10,000M_{\odot}$ black hole will always be classified as a star-forming galaxy at an ionization parameter of $\log{U}=-2$ unless the Eddington ratio falls below 0.1, even without including the effects of photoionization from stars.  At Eddington ratios below 0.1, the AGN will be less luminous and the line ratios will likely be dominated by photoionization from star formation, suggesting that accreting black holes in this mass range may rarely be detected as AGNs using this standard optical diagnostic diagram. As \citet{greene2007} find a typical Eddington ratio for low mass black holes of ~0.4, our results leave open the possibility that the use of these diagnostics to confirm IMBH candidates below $\approx 10^4M_{\odot}$ could severely bias the inferred low-mass black hole occupation fraction. We note that apart from the limitations of these narrow line region diagnostics, \citet{chakravorty2014} have shown that such low mass black holes may not even show a broad line region and \citet{baldassare2016} have shown that, in low mass galaxies, broad lines can actually be due to supernovae, further emphasizing the shortcomings of optical diagnostics in finding AGNs powered by IMBHs.
\subsection{Detectability of IMBHs with Current Facilities}

Our results have important consequences for high spatial resolution optical spectroscopic follow-up studies of accreting IMBH candidates. While it is well-known that dilution from circumnuclear star formation significantly limits the diagnostic power of optical spectroscopy in identifying AGNs in low mass galaxies using large aperture surveys such as SDSS, it has been assumed that if contamination from star formation is reduced using high spatial resolution optical spectroscopy, the AGN can be revealed. To illustrate this point, the typical [O~III] luminosities of star forming dwarf galaxies within the $3\arcsec$ SDSS fiber is about $10^{39}~erg/s$ \citep{reines2013}, which corresponds approximately to the Eddington limit for a $1000M_{\odot}$ black hole, given a conversion between $L_{bol}$ and $L_{[O~III]}$ of about $100$ for low luminosities \citep[e.g.,][]{lamastra2009}. Star formation would therefore dominate over the AGN emission from a $10^3-10^4M_{\odot}$ black hole for all but the most highly accreting systems. However, assuming that the star formation rate is reasonably uniform within the $3\arcsec$ aperture, star formation within the $0.2\arcsec$ spaxels of an IFU with AO would only contribute about $10^{36}~erg/s$ to the [O~III] emission, allowing an accreting IMBH as low as $100-1000M_{\odot}$ to be detected, \textit{if} they exhibited line ratios similar to higher mass black holes. Our results demonstrate that such follow-up optical spectroscopic studies will misidentify accreting IMBHs, which will masquerade as star forming galaxies, \textit{even when the effects of contamination from surrounding star formation are removed}.

\begin{figure}
\includegraphics[width=0.495\textwidth]{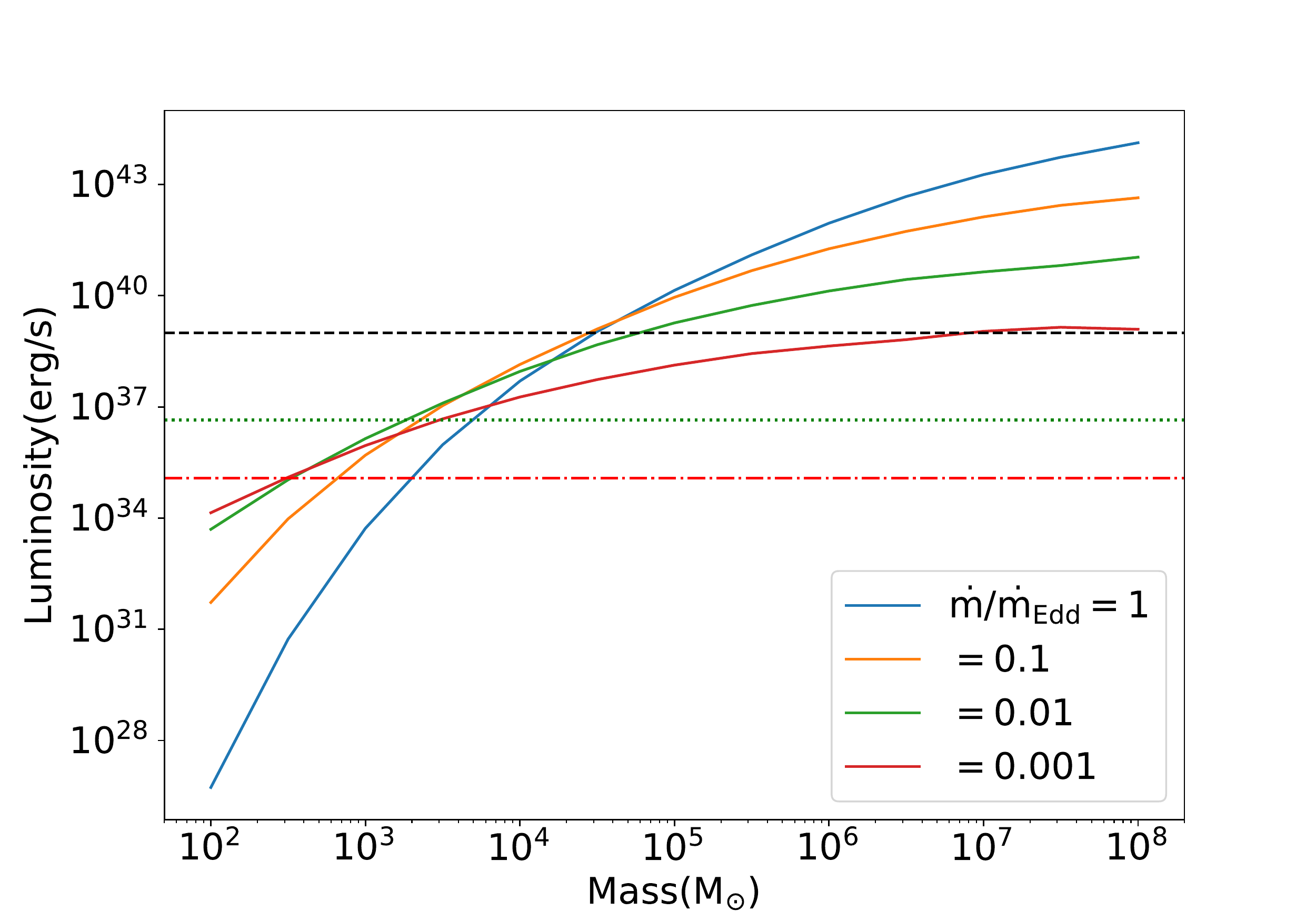}
\caption{Luminosity of [O~III] as a function of black hole mass for the $n_H=300~cm^{-3}$, solar metallicity, $\log{U}=-3$ model, for a range of Eddington ratios.  Also plotted are the median [O~III] luminosity of SDSS star forming dwarf galaxies (black dotted line), the aperture-reduced [O~III] luminosity from star formation detected in a $0.2\arcsec$ IFU spaxel (green dotted line), and the [O~III] luminosity corresponding to a detection threshold of $10^{-17}~erg/cm^2/s$ (red dot-dashed line).  Note that the horizontal luminosity thresholds displayed correspond to a distance of 10 Mpc and even the lowest black hole masses could be detected in closer sources using an IFU.}
\label{fig6}
\end{figure}

Using our black hole mass dependent models, we plot in Figure \ref{fig6} the [O~III] luminosity as a function of black hole mass for a wide range of Eddington ratios. While [N~II] is generally the weaker line, and the limiting factor in identifying an AGN, we have chosen to show [O~III] luminosity, as the [O~III]/H$\beta$ line ratio showed a wider range of values across our models.  We also show the median [O~III] luminosity of star forming dwarf galaxies ($\log M_* < 9.5$) from the SDSS $3\arcsec$ fiber (black dotted line), the aperture-reduced [O~III] luminosity from star formation in a $0.2\arcsec$ IFU spaxel (green dotted line), and the [O~III] line luminosity assuming a detection threshold of $10^{-17}~erg/cm^2/s$ assuming a 10 Mpc distance. As can be seen, using our mass dependent models and assuming a distance of 10 Mpc,  the [O~III] luminosity of black holes with masses down to $10^3M_{\odot}$ could be identified. However, our work shows that such IMBHs would  typically not
be identified as AGN using the widely-used BPT
classifications for higher mass black holes.

\subsection{Additional Considerations and Caveats}
Our goal in this paper is to examine the first order effects of black hole mass on the widely used BPT line ratios.  We have chosen a simple accretion disk, together with a power law and soft excess component, to model the AGN continuum. The predicted line ratios will change with more complex models. However, the purpose of this investigation is to explore the dependence of predicted emission line strengths on the SED of the accretion disk as black hole mass varies.    

We note emission line strengths can also be affected by many other physical processes not included in this initial study.  This model does not include the effects of shocks generated by AGN outflows or starburst driven winds, which can alter the emission line spectrum \citep{allen2008, kewley2013}, and can even generate emission lines consistent with star-forming galaxies in pure AGN models \citep{molina2018}.  Our model also takes into account ionizing radiation from the AGN only, however, as discussed in the introduction, dilution from star formation only exacerbates the effects discussed here.  The limitations of optical diagnostics are even more severe in low metallicity galaxies at all black hole masses \citep{groves2006}.

In addition, our models have assumed a simple geometrically thin, optically thick disk. This assumes the accretion is radiatively efficient, and viscous heating is balanced by radiative cooling. When the mass accretion rate falls below $0.01$ $\dot{m}/\dot{m}_{Edd}$, the accretion flow is predicted to be advection-dominated and radiatively inefficient (RIAF) \citep{narayan1998, ho2008, yuan2014b}, producing a significantly different SED that may lack a standard big blue bump with much of the emission arising in the IR \citep{quataert2000}.  We also note that we do not take into account the effect of black hole spin, which would affect the innermost stable orbit, and therefore the temperature of the accretion disk, which in turn would affect the shape of the emergent SED. Our model also does not take into account radiative transfer through the atmosphere of the disk. \par

\section{Conclusions} \label{sec:conclusions}
In this work, we have modeled the BPT emission lines from an AGN and explored for the first time the dependence of these line ratios on black hole mass over the range $10^2M_{\odot}-10^8M_{\odot}$, as well as Eddington ratio, ionization parameter $U$, metallicity, and number density $n_H$.  Our photoionization models assume purely an AGN ionizing continuum, with a standard geometrically thin, optically thick accretion disk, a power law component, and a soft excess.  Based on this model, we have demonstrated that the standard optical spectral classification schemes used to identify higher mass black holes do not apply when the black hole mass falls below $\approx10^4M_{\odot}$.  These IMBHs will fall outside the widely adopted AGN regime for a significant range of parameter space compared with more massive black holes.  Note that there are significant uncertainties in this work due to the currently unknown nature of the intrinsic SED of these objects and this paper explores the mass dependence assuming a simplified accretion disk model.  This important result, however, demonstrates that, \textit{independent} of the effects of dilution from star formation in dwarf galaxies, BPT diagnostics will be highly incomplete in confirming accreting IMBHs. Given this limitation, another promising tool to hunt for IMBHs is the use of infrared coronal lines. The power of these diagnostics in finding AGNs in the low mass regime has been demonstrated by Spitzer \citep{satyapal2007, satyapal2008, satyapal2009}, and the promise of the \textit{James Webb Space Telescope} in finding new populations is discussed in \citet{cann2018}.

\acknowledgments

We thank the referee for their thoughtful comments and insight, particularly regarding the use of mythical IFUs. J.M.C. gratefully acknowledges support from an NSF GRFP and a Mason 4-VA Innovation grant.  L.B. acknowledges support from a National Science Foundation grant (AST-1715413). C.S.R. thanks the UK Science and Technology Facilities Council for support under Consolidated Grant ST/R000867/1.

%

\vspace{5mm}


\software{astropy \citep{astropy2013},  
          Cloudy \citep{ferland2017}, 
          TOPCAT \citep{taylor2005}
          }

\end{document}